\documentclass[a4paper,twocolumn]{revtex4}

\usepackage{amsmath,amssymb,amsfonts}
\usepackage{graphicx}

\newcommand{\ham}{\mathcal{H}}

\begin{document}

\title{Revisiting waterlike network-forming lattice models}

\author{M.~Pretti}

\affiliation{Dipartimento di Fisica and CNISM, Politecnico di
Torino, Corso Duca degli Abruzzi 24, I-10129 Torino, Italy}

\affiliation{Center for Statistical Mechanics and Complexity,
CNR-INFM Roma 1, Piazzale Aldo Moro 2, I-00185 Roma, Italy}

\author{C.~Buzano}

\author{E.~{De~Stefanis}}

\affiliation{Dipartimento di Fisica and CNISM, Politecnico di
Torino, Corso Duca degli Abruzzi 24, I-10129 Torino, Italy}


\begin{abstract}
In a previous paper [J. Chem. Phys. \textbf{129}, 024506 (2008)]
we studied a 3-dimensional lattice model of a network-forming
fluid, recently proposed in order to investigate water anomalies.
Our semi-analytical calculation, based on a cluster-variation
technique, turned out to reproduce almost quantitatively several
Monte Carlo results and allowed us to clarify the structure of the
phase diagram, including different kinds of orientationally
ordered phases. Here, we extend the calculation to different
parameter values and to other similar models, known in the
literature. We observe that analogous ordered phases occur in all
these models. Moreover, we show that certain ``waterlike''
thermodynamic anomalies, claimed by previous studies, are indeed
artifacts of a homogeneity assumption made in the analytical
treatment. We argue that such a difficulty is common to a whole
class of lattice models for water, and suggest a possible way to
overcome the problem.

\end{abstract}


\maketitle

\section{Introduction}

Water is extremely abundant in nature, and it is of enormous
importance from a biological, technological as well as
environmental point of view, because of its various anomalous
properties~\cite{CabaneVuilleumier2005}. For instance, it is well
known that liquid water exhibits unusually large heat capacity and
dielectric constant. Furthermore, at ordinary pressures, the solid
phase (ice) is less dense than the corresponding liquid phase,
while the latter displays a temperature of maximum density,
slightly above the freezing transition. In spite of great research
efforts on
water~\cite{EisenbergKauzmann1969,Franks1982,Stanley2003}, a fully
consistent theory of such (and many other) anomalies from first
principles is not yet available. Nonetheless, it is well
established that most anomalies are related to the ability of
water molecules to form a network of hydrogen bonds.

Following this view, so-called network-forming fluids have been
the subject of several theoretical investigations. Among different
possible approaches, a number of studies have been developed in
the framework of simplified lattice
models~\cite{BellLavisI1970,BellLavisII1970,SouthernLavis1980,HuckabyHanna1987,PatrykiejewPizioSokolowski1999,BuzanoDestefanisPelizzolaPretti2004,Bell1972,WilsonBell1978,WhitehouseChristouNicholsonParsonage1984,BellSalt1976,MeijerKikuchiPapon1981,MeijerKikuchiVanRoyen1982,LavisSouthern1984,BesselingLyklema1994,BesselingLyklema1997,RobertsDebenedetti1996,RobertsPanagiotopoulosDebenedetti1996,RobertsKarayiannakisDebenedetti1998,PrettiBuzano2004,PrettiBuzano2005,GirardiBalladaresHenriquesBarbosa2007,GirardiSzortykaBarbosa2007,BuzanoDeStefanisPretti2008,SastrySciortinoStanley1993jcp,BorickDebenedettiSastry1995,SastryDebenedettiSciortinoStanley1996,RebeloDebenedettiSastry1998}.
Various types of model molecules with orientation-dependent
interactions have been used, in either
two~\cite{BellLavisI1970,BellLavisII1970,SouthernLavis1980,HuckabyHanna1987,PatrykiejewPizioSokolowski1999,BuzanoDestefanisPelizzolaPretti2004}
or
three~\cite{Bell1972,WilsonBell1978,WhitehouseChristouNicholsonParsonage1984,BellSalt1976,MeijerKikuchiPapon1981,MeijerKikuchiVanRoyen1982,LavisSouthern1984,BesselingLyklema1994,BesselingLyklema1997,RobertsDebenedetti1996,RobertsPanagiotopoulosDebenedetti1996,RobertsKarayiannakisDebenedetti1998,PrettiBuzano2004,PrettiBuzano2005,GirardiBalladaresHenriquesBarbosa2007,GirardiSzortykaBarbosa2007,BuzanoDeStefanisPretti2008,SastrySciortinoStanley1993jcp,BorickDebenedettiSastry1995,SastryDebenedettiSciortinoStanley1996,RebeloDebenedettiSastry1998}
dimensions. A natural choice for water is a 3-dimensional model
molecule with four bonding arms arranged in a tetrahedral
symmetry~\cite{Bell1972,WilsonBell1978,WhitehouseChristouNicholsonParsonage1984,BellSalt1976,MeijerKikuchiPapon1981,MeijerKikuchiVanRoyen1982,LavisSouthern1984,BesselingLyklema1994,BesselingLyklema1997,RobertsDebenedetti1996,RobertsPanagiotopoulosDebenedetti1996,RobertsKarayiannakisDebenedetti1998,PrettiBuzano2004,PrettiBuzano2005,GirardiBalladaresHenriquesBarbosa2007,GirardiSzortykaBarbosa2007,BuzanoDeStefanisPretti2008}.
Two arms represent the hydrogen (H) atoms, which are positively
charged and act as donors for the H bond, whereas the other two
represent the negatively charged regions of the
$\mathrm{H}_2\mathrm{O}$ molecule (``lone pairs''), acting as
H-bond acceptors. As far as the lattice is concerned, the
body-centered cubic (bcc) lattice is suitable for the tetrahedral
molecule, as the latter can point its arms toward four out of
eight nearest neighbors of each given site. The above features are
common to several models, which differ in the form of interactions
and in the set of allowed configurations.

In the early model proposed by
Bell~\cite{Bell1972,WilsonBell1978,WhitehouseChristouNicholsonParsonage1984},
molecules can point their arms only toward nearest neighbor sites.
An attractive energy is assigned to every pair of occupied nearest
neighbors, with an extra contribution if a H bond is formed, i.e.,
if a donor arm points toward an acceptor arm. Moreover, a
repulsive energy is assigned to certain triplets of occupied
sites, in order to account for the difficulty of forming H bonds
by closely-packed water molecules. Minor variations of this model
have also been investigated: Bell and Salt~\cite{BellSalt1976},
and subsequently Meijer and
coworkers~\cite{MeijerKikuchiPapon1981,MeijerKikuchiVanRoyen1982},
have replaced the three-body interaction with a simple
next-nearest-neighbor repulsion, whereas Lavis and
Southern~\cite{LavisSouthern1984} have defined a simplified model
with no distinction between donors and acceptors. All these
studies predict a liquid-vapor coexistence and two different
low-temperature phases, characterized by orientational order and
different densities. At zero temperature, the low-density phase is
an ideal diamond network made up of H-bonded water molecules, with
half the lattice sites left empty, resembling the structure of ice
Ic (cubic ice). Conversely, the high-density phase is made up of
two interpenetrating diamond structures, with all the sites
occupied, resembling the structure of ice VII (a high pressure
form of ice). If the models retain bond asymmetry, i.e., donors
and acceptors are distinguished, then both zero-temperature phases
possess a residual entropy.

Debenedetti and coworkers have studied a similar
model~\cite{RobertsDebenedetti1996,RobertsPanagiotopoulosDebenedetti1996,RobertsKarayiannakisDebenedetti1998},
in which water molecules have an extra number of nonbonding
configurations and the close-packing energy penalty occurs only
when two molecules in a triplet form a H bond. At zero
temperature, this model still predicts the two ordered H-bond
networks of the Bell model. At finite temperature, these (icelike)
phases have not been investigated, as the cited works were mainly
focused on metastable liquid water. Indeed, the model seems to
support the popular ``second critical point''
conjecture~\cite{Harrington1997,MishimaStanley1998}, originally
proposed by Stanley and
coworkers~\cite{PooleSciortinoEssmannStanley1992}. Two authors of
the present paper have also studied a simplified version of this
model, without the donor-acceptor
asymmetry~\cite{PrettiBuzano2004,PrettiBuzano2005}.

A variation of the Bell model has been considered by Besseling and
Lyklema~\cite{BesselingLyklema1994,BesselingLyklema1997}, who have
taken into account only nearest-neighbor interactions, namely, an
attractive term for H-bonded molecules and a repulsive term for
nonbonded molecules. Even in this case, the two different ordered
phases described above are stable at zero temperature, but they
have not been investigated at finite temperature. The cited
studies were indeed devoted to liquid-vapor interface
properties~\cite{BesselingLyklema1994} and to hydrophobic
hydration thermodynamics~\cite{BesselingLyklema1997}, for which
the authors found good agreement with experiments. These results
were obtained in the so-called quasi-chemical or Bethe
approximation, i.e., a first-order approximation which takes into
account correlations over clusters made up of two nearest-neighbor
sites.

Girardi and coworkers~\cite{GirardiBalladaresHenriquesBarbosa2007}
have recently performed extensive Monte Carlo simulations for the
above model, in the simplified version with symmetric bonds. This
work claims the onset of two liquid phases of different densities,
which can coexist in equilibrium, in agreement with Stanley's
conjecture~\cite{PooleSciortinoEssmannStanley1992}. The
coexistence line seems to terminate in a critical point, while the
lower-density phase exhibits a temperature of maximum density,
depending on pressure.

In our previous paper~\cite{BuzanoDeStefanisPretti2008} we
analyzed the latter model by a generalized first-order
approximation, based on a four-site tetrahedral cluster. Such
semi-analytical calculation turns out to reproduce, with
remarkable accuracy, most physical properties obtained by
simulations. Nevertheless, the resulting phase diagram is quite
different from the one proposed in the original
paper~\cite{GirardiBalladaresHenriquesBarbosa2007}. The two
different condensed phases exhibit orientational order, which
suggests to identify them as a temperature evolution of the two
zero-temperature network structures introduced above. Furthermore,
the two claimed critical points turn out to be in fact
tricritical, and two critical lines appear, related to two
different kinds of symmetry breaking. In conclusion, the phase
diagram is more complex and richer than expected, but
unfortunately quite far from the real water phase diagram.

In the current paper, we exploit the effectiveness of our
approximation scheme, with a twofold purpose. On the one hand, we
analyse in more detail the simple model by Girardi and coworkers
(GBHB), exploring the phase diagram for different parameter
values. For strong repulsive interaction (slightly weaker than the
H~bond), we obtain an even richer phase diagram, including a new
ordered phase, appearing at finite temperature. On the other hand,
we reconsider more complex models with asymmetric bonds, such as
the Besseling-Lyklema (BL) model and the original Bell model.

Several investigations devoted to this kind of models have
hypothesized a fully homogeneous (i.e., site-independent)
probability distribution of molecular configurations, thus
excluding the possibility of orientational ordering of the type
observed in the GBHB model~\cite{BuzanoDeStefanisPretti2008}.
Conversely, we find out that analogous ordered phases occur even
in the BL and Bell models. The liquid phase reported by previous
papers~\cite{Bell1972,BesselingLyklema1994} (displaying waterlike
density anomalies) turns out to be an artifact of the
aforementioned homogeneity assumption. Furthermore, density
anomalies appear in the lower-density ordered phase, which makes
it questionable to regard this phase as a representation of some
real ice form. Let us note that density anomalies in an icelike
phase have been previously noticed by Meijer and
coworkers~\cite{MeijerKikuchiVanRoyen1982}, investigating the
Bell-Salt model~\cite{BellSalt1976}. We discuss the reasons
underlying these difficulties, which we argue to be common to a
wide class of network-forming lattice models for water, and
finally suggest a possible way to circumvent the problem.

The paper is organized as follows. In Section~II we introduce a
generic model hamiltonian incorporating the three models under
investigation. In Section~III we describe in more detail the
ground states of the models. In Section~IV we explain the
cluster-variation technique employed for the calculation.
Section~V reports the results, and a comparison to previous
semi-analytical~\cite{BesselingLyklema1994} and
numerical~\cite{WhitehouseChristouNicholsonParsonage1984}
investigations. Section~VI contains the conclusions.

\section{The model hamiltonian}

\begin{figure}[t]
  \includegraphics*[60mm,107mm][140mm,182mm]{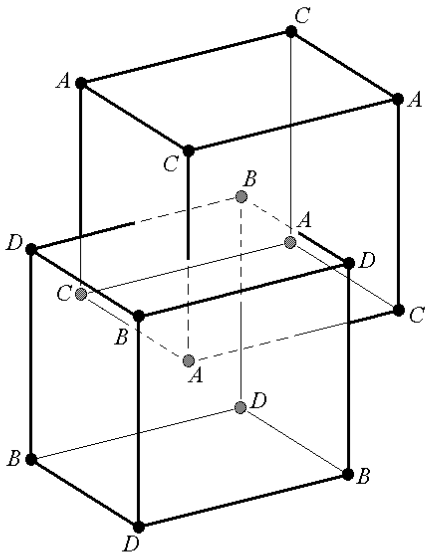}
  \caption
  {
    Two conventional (cubic) cells of the body centered cubic (bcc) lattice.
    $A,B,C,D$ denote four interpenetrating face-centered cubic (fcc) sublattices.
  }
  \label{fig:reticolo}
\end{figure}

As mentioned in the Introduction, we study three different models
of increasing complexity (GBHB, BL, and Bell), defined on a bcc
lattice (Fig.~\ref{fig:reticolo}).
It is possible to write a single hamiltonian incorporating the
three models. For all the models, a lattice site can be empty or
occupied by a molecule having a tetrahedral structure (4 bonding
arms, which can point toward 4 out of 8 nearest neighbors of each
given site).

\begin{figure}[t]
  \includegraphics*[60mm,107mm][140mm,182mm]{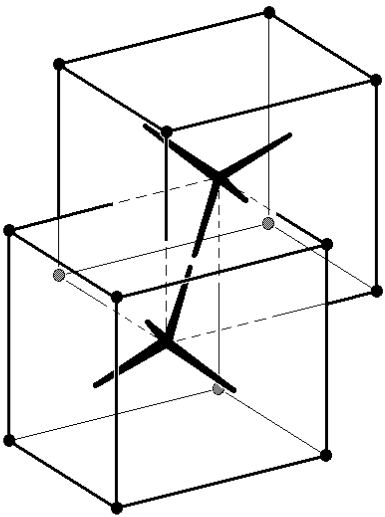}
  \caption
  {
    Two model molecules in a bonding configuration.
    The arm configuration is $i=1$ for the lower molecule
    and $i=2$ for the upper one (see the text).
  }
  \label{fig:molecole}
\end{figure}

In the simplest case (GBHB model), a hydrogen bond is formed,
yielding an attractive energy $-\eta<0$, whenever two
nearest-neighbor molecules point an arm toward each other, with no
distinction between donor and acceptor. Each molecule can thus
have only two different configurations, which we simply denote as
$1$ and $2$ (see Fig.~\ref{fig:molecole}).
Moreover, a repulsive energy $-\epsilon>0$ is assigned to every
pair of nearest-neighbor sites occupied by water molecules. This
term is meant to penalize neighbor molecules not forming H bonds.
Finally, a chemical potential contribution $-\mu$ is taken into
account for every occupied site (in this paper we always consider
a grand-canonical description).

In the BL model~\cite{BesselingLyklema1994}, the above picture is
a little more complex. The bonding arms of each molecule are
``decorated'' by a sign (two ``$+$'' and two ``$-$''),
respectively denoting donors and acceptors. A H bond is formed
only if the two bonding arms, pointing toward each other, carry
one ``$+$'' sign and one ``$-$'' sign, or vice versa. As a
consequence, a water molecule can assume 6 distinguishable ``sign
configurations'' for each of the 2 ``arm configurations''
described above, adding up to 12 different configurations.

The Bell model~\cite{Bell1972} is equivalent to the BL model in
the bonding mechanism, but incorporates different
orientation-independent interactions. The two-molecule interaction
is attractive ($-\epsilon<0$), whereas a repulsive energy
$-\gamma>0$ is assigned to any (fully occupied) 3-site cluster,
made up of two second neighbors plus one of their common first
neighbors.

We can write the general model hamiltonian as
\begin{eqnarray}
  \ham & = &
  - \epsilon \sum_{(r,s)} n_{i_{r}} n_{i_{s}}
  - \mu \sum_r n_{i_{r}}
  \label{eq:hamiltonian}
  \\ &&
  - \eta \sum_{(r,s)} h_{i_{r}i_{s}}
  - \gamma \sum_{(r,s,t)} n_{i_{r}} n_{i_{s}} n_{i_{t}}
  \nonumber
  ,
\end{eqnarray}
where $r,s,t$ denote lattice sites, $i_{r},i_{s},i_{t}$ denote
their respective configurations, $n_i$ is an ``occupation
function'', defined as $n_i=0$ if $i=0$ (empty site), $n_i=1$
otherwise (occupied site), and $h_{ij}$ is a ``bond function'',
defined as $h_{ij}=1$ if the pair configuration $(i,j)$ represents
a H~bond, and $h_{ij}=0$ otherwise. The summations denoted by $r$,
$(r,s)$, and $(r,s,t)$ respectively run over sites, nearest
neighbor pairs, and the 3-site clusters defined above. Let us note
that the first line of Eq.~\eqref{eq:hamiltonian} is formally
equivalent to a lattice-gas hamiltonian, although the GBHB and the
BL models are characterized by a repulsive interaction energy
($\epsilon<0$). The first term of the second line represents the
H-bond energy, which is always attractive ($\eta>0$), whereas the
last term takes into account the 3-body repulsive interaction
($\gamma<0$). As mentioned above, the latter occurs only in the
Bell model, whereas the GBHB and BL models are characterized by
$\gamma=0$. Concerning configuration indices, we can always denote
a vacancy (empty site) by $i=0$, but molecular configurations need
to be defined in different ways, depending on whether the model
assumes asymmetric bonds (donor-acceptor distinction) or symmetric
bonds (no distinction). In the latter case (GBHB model), the two
possible arm configurations, which we respectively denote by
$i=1,2$ (see Fig.~\ref{fig:molecole}), provide a full description
of the molecule configurations. In the other case we have to
understand that $i$ is in fact a multi-index $(i,i')$, where $i'$
denotes the sign configuration. As a consequence, in general the
precise definition of the bond function $h_{ij}$ turns out to
depend on the spatial orientation of the lattice bond considered.
It can be made unambiguous in the symmetric bond case, by defining
a special order for the sites in each nearest neighbor pair. Such
order may be for instance the one indicated by the arms of a
molecule in the $i=1$ configuration, in which case we obtain the
following simple definition: $h_{ij}=1$ if $i=1$ and $j=2$, and
$h_{ij}=0$ otherwise.

\section{Ground states}

The models under investigation can exhibit two different ground
states with different densities, in which the bonding arms of the
molecules form ordered network structures, as described in the
Introduction. Such ordered structures naturally split the bcc
lattice into four (mutually exclusive) interpenetrating fcc
sublattices, which we denote by $A,B,C,D$ (see
Fig.~\ref{fig:reticolo}). Assuming here that $i,j,k,l$ denote the
arm configurations of all sites placed on the $A,B,C,D$
sublattices, respectively, one can see that the low density
structure (single diamond network) is fourfold degenerate and can
be represented by the four alternative sublattice configurations
$(i,j,k,l)=(1,2,0,0), (0,1,2,0), (0,0,1,2), (2,0,0,1)$ (which can
be obtained from one another by a circular permutation). In each
structure, two sublattices (respectively, $AB$, $BC$, $CD$, and
$DA$) are occupied, while the other two sublattices are empty.
Conversely, the high-density structure (two intertwined diamond
networks) turns out to be twofold degenerate and can be
represented by the two alternative sublattice configurations
$(i,j,k,l)=(1,2,1,2), (2,1,2,1)$. Both these configurations have
all lattice sites occupied, but they differ in the pairs of bonded
sublattices, which is, $AB$ and $CD$, in the former case, or
alternatively $BC$ and $DA$ in the latter.

In this paper, we do not investigate in detail the ground state
phase diagrams, which depend on the interaction parameters of each
different model. Let us only remark an important difference
between the simpler GBHB model, with symmetric bonds, and the
other two models with asymmetric bonds. In the former case,
molecule configurations are fully defined by their arm
configurations, so that the formation of a diamond network
immediately implies that all molecules are maximally bonded (each
molecule can participate in four bonds at most). Conversely, in
the latter case, molecules still have the freedom of arranging
their sign configurations in order to respect the ice rule (i.e.,
to realize the correct donor-acceptor pairings), giving rise to an
exponential number of global configurations with equal energy, and
therefore to a residual ground-state entropy. An approximate
calculation of this entropy was first performed in 1935 by
Pauling, who obtained a value of $\ln(3/2)$ per
molecule~\cite{Pauling1935}. This value turns out to be in
extremely good agreement with both experimental measurements and
simulations (see Ref.~\cite{BergYang2007} and references therein).

\section{Cluster-variation approximation}

\begin{figure}[t]
  \resizebox{80mm}{!}{\includegraphics*[75mm,130mm][155mm,160mm]{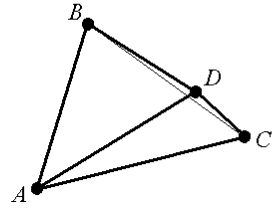}}
  \caption
  {
    Tetrahedral cluster, made up of four sites placed
    on the four different $A,B,C,D$ sublattices.
    $AB$, $BC$, $CD$, and $DA$ are nearest-neighbor pairs;
    $AC$ and $BD$ are next-nearest-neighbor pairs.
  }
  \label{fig:tetraedro}
\end{figure}

The ground state structures described above suggest to write the
average (grand-canonical) energy per site in the following form
\begin{equation}
  w = \sum_{i,j,k,l}  p_{ijkl} \ham_{ijkl} ,
  \label{eq:intenergy}
\end{equation}
where $p_{ijkl}$ denotes the joint probability distribution for
the configurations of a tetrahedral cluster like the one depicted
in Fig.~\ref{fig:tetraedro}, whose sites are placed respectively
on the four different $A,B,C,D$ sublattices, and $\ham_{ijkl}$ is
a suitable function, which we denote as tetrahedron hamiltonian.
The sum is understood to run over all possible configurations of
the four sites. By grand-canonical energy, we mean $w=u-\mu\rho$,
where $u$ is the internal energy per site and $\rho$ is the
density, i.e., the average occupation probability
\begin{equation}
  \rho = \sum_{i,j,k,l}  p_{ijkl} \frac{n_i+n_j+n_k+n_l}{4} .
  \label{eq:density}
\end{equation}
At first, let us consider the model with symmetric bonds (GBHB
model), so that the configuration indices denote only arm
configurations. It is easy to verify that the tetrahedron
hamiltonian can be written as follows,
\begin{equation}
  \ham_{ijkl} = \ham_{ijk} + \ham_{jkl} + \ham_{kli} + \ham_{lij}
  ,
  \label{eq:tetraham}
\end{equation}
where
\begin{equation}
  \ham_{ijk} = - \epsilon \, n_i n_j - \mu \, n_i /4
  - \eta \, h_{ij} - 3 \gamma \, n_i n_j n_k
  .
  \label{eq:triham}
\end{equation}
Eq.~\eqref{eq:tetraham} shows that the tetrahedron hamiltonian
turns out to be invariant under circular permutation of the
configuration indices $i,j,k,l$. As far as Eq.~\eqref{eq:triham}
is concerned, numerical coefficients are meant to adjust counting
of the different terms. The one-site (chemical potential) term is
divided by 4, which is, the number of sites in the tetrahedron.
The two-site terms have a unit coefficient, because there are
exactly 4 nearest-neighbor pairs in the tetrahedron and 4
nearest-neighbor pairs per site in the lattice. Finally, the
three-site term is multiplied by 3, because there are 4 triangle
clusters in the tetrahedron, but 12 triangle clusters per site in
the lattice.

Let us note that the meaning of Eq.~\eqref{eq:intenergy}, and of
the related definition of the tetrahedron hamiltonian, is that
every elementary tetrahedron is assumed to give on average the
same contribution to the energy of the system. In other words, we
assume that possible breaking of translational invariance in the
thermodynamic state of the system can only occur at the level of
the four sublattices identified by the ground states, whereas the
thermodynamic state of each individual sublattice is assumed to
remain translationally invariant. This assumption is supported by
Monte Carlo simulations, performed for the GBHB
model~\cite{BuzanoDeStefanisPretti2008}.

As far as entropy is concerned, we make use of the very same
approximation employed for the GBHB
model~\cite{BuzanoDeStefanisPretti2008}. In the cited paper, we
deduced the approximation as an instance of Kikuchi's
cluster-variation method~\cite{Kikuchi1951,An1988,Pelizzola2005}.
The latter is a generalized mean-field theory, which describes
correlations up to the size of certain maximal clusters. In
general, one obtains a free-energy functional in the cluster
probability distributions, to be minimized, according to the
variational principle of statistical mechanics. Here, we derive
the variational free energy, according to a heuristic argument
analogous to the one originally suggested by
Guggenheim~\cite{Guggenheim1935}. One assumes that the entropy per
site can be evaluated as a difference of two terms. The former is
the information entropy associated with the 4-site probability
distribution $p_{ijkl}$, which takes into account correlations
among 4 configuration variables on a tetrahedral cluster (i.e., on
the 4 different sublattices). The latter can be viewed as a
correction term, which ensures that, if the tetrahedron
distribution factorizes into a product of single-site
probabilities, the mean-field (Bragg-Williams) entropy
approximation is recovered.

\begin{figure}[t]
  \includegraphics*[60mm,107mm][140mm,157mm]{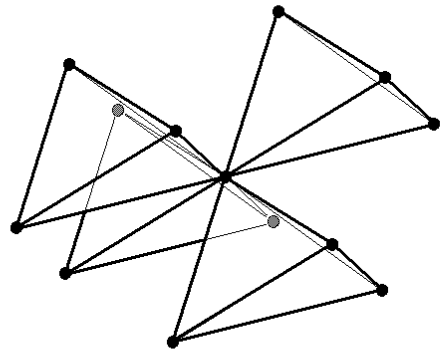}
  \caption
  {
    Local structure of a Husimi lattice,
    made up of tetrahedral blocks.
  }
  \label{fig:cactus}
\end{figure}

The variational grand-canonical free energy per site $\omega = w -
Ts$ ($T$ being the temperature, expressed in energy units, and $s$
the entropy per site, in natural units) can be finally written as
\begin{equation}
  \frac{\omega}{T} =
  \sum_{i,j,k,l} p_{ijkl}
  \left[ \frac{\ham_{ijkl}}{T} + \ln p_{ijkl}
  - \frac{3}{4} \ln \left( p^A_i p^B_j p^C_k p^D_l \right) \right] ,
  \label{eq:func}
\end{equation}
where $p^X_i$ is the probability of the $i$~configuration for a
site on the $X$~sublattice ($X=A,B,C,D$). These probabilities are
of course defined as marginals of the tetrahedron distribution as
\begin{align}
  p^A_i & = \sum_{j,k,l} p_{ijkl},
  &
  p^B_j & = \sum_{k,l,i} p_{ijkl},
  \nonumber \\
  p^C_k & = \sum_{l,i,j} p_{ijkl},
  &
  p^D_l & = \sum_{i,j,k} p_{ijkl}.
  \label{eq:marginals}
\end{align}
The variational free energy in Eq.~\eqref{eq:func} is thus a
function of the only tetrahedron distribution $p_{ijkl}$. Let us
note that such a free energy is also sometimes referred to as
(generalized) first-order approximation (on the tetrahedron
cluster), and turns out to be exact for a suitable Husimi
lattice~\cite{Pretti2003,Lage-CastellanosMulet2008}, with the
(tetrahedral) building blocks arranged as in
Fig.~\ref{fig:cactus}.

The free energy minimization, taking into account the
normalization constraint
\begin{equation}
  \sum_{i,j,k,l} p_{ijkl} = 1 ,
  \label{eq:constraint}
\end{equation}
can be performed by the Lagrange multiplier method, yielding
\begin{equation}
  p_{ijkl} = \zeta^{-1} e^{-\ham_{ijkl}/T}
  \left( p^A_i p^B_j p^C_k p^D_l  \right)^{3/4} ,
  \label{eq:cvmeq}
\end{equation}
where $\zeta$ is related to the Lagrange multiplier, and can be
determined by imposing the constraint Eq.~\eqref{eq:constraint}.
One obtains
\begin{equation}
  \zeta =
  \sum_{i,j,k,l} e^{-\ham_{ijkl}/T}
  \left( p^A_i p^B_j p^C_k p^D_l \right)^{3/4}
  .
  \label{eq:costnorm}
\end{equation}

Eq.~\eqref{eq:cvmeq}, together with Eqs.~\eqref{eq:costnorm}
and~\eqref{eq:marginals}, provides a fixed-point equation for the
tetrahedron distribution $p_{ijkl}$, which can be solved
numerically by simple iteration (natural iteration
method~\cite{Kikuchi1974}). This numerical procedure can be proved
to lower the free energy at each
iteration~\cite{Kikuchi1974,Pretti2003}.

From the tetrahedron distribution, one can determine the thermal
average of every observable. The density can be computed by
Eq.~\eqref{eq:density}, the internal energy $u=w+\mu\rho$ by
Eqs.~\eqref{eq:intenergy} and~\eqref{eq:density}, and the free
energy by Eq.~\eqref{eq:func}. The latter can also be related to
the normalization constant as
$\omega=-T\ln\zeta$~\cite{Pretti2003}, so that the entropy
reads~$s=\ln\zeta+w/T$. Finally, assuming the volume per site
equal to unity, pressure can be determined, in energy units, as
$P=-\omega$. In the presence of multiple solutions, i.e., of
competing phases, the thermodynamically stable one is selected by
the lowest free energy (highest pressure) value.

As far as first order transitions are concerned, they can be
easily determined by finding (numerically) a change of sign in the
difference between the free energy values of two minima of the
variational free energy. Second order (critical) transitions are
numerically more delicate, since they are characterized by a free
energy minimum becoming a saddle point. In this case, it is
convenient to determine changes of sign in some eigenvalue of the
Hessian matrix of the variational free energy. The elements of the
latter can be written as
\begin{eqnarray}
  && \frac{1}{T} \, \frac{\partial^2 \omega}{\partial p_{ijkl} \partial \, p_{i'j'k'l'}}
  = \label{eq:hessian} \\ &&
  = \frac{\delta_{ii'}\delta_{jj'}\delta_{kk'}\delta_{ll'}}{p_{ijkl}}
  - \frac{3}{4} \left( \frac{\delta_{ii'}}{p_i^A} + \frac{\delta_{jj'}}{p_j^B}
  + \frac{\delta_{kk'}}{p_k^C} + \frac{\delta_{ll'}}{p_l^D} \right)
  \nonumber
  ,
\end{eqnarray}
where $\delta$ denotes Kronecker delta.

As mentioned above, these calculations are valid in the case of
symmetric bonds (GBHB model), but they have to be generalized, in
order to deal with the asymmetric-bond models (BL and Bell
models). In the latter cases one has to take into account both arm
and sign configurations. We make the simplifying assumption that
all sign configurations are equally probable, for a molecule in a
given arm configuration. Such assumption is reasonable, since it
turns out to lead to a very good approximation for the zero-point
entropy of perfect ice (where correlations among sign
configurations are expected to be maximal). Indeed, in
Ref.~\onlinecite{BesselingLyklema1994}, Besseling and Lyklema
showed that Pauling's calculation is exactly equivalent to the
quasi-chemical (Bethe) approximation, with the extra assumption of
equally probable sign configurations. A simple, though quite
tedious, calculation proves that the same result is recovered by
our cluster approximation. In a few words, this is related to the
fact that each tetrahedral (4-site) cluster contains only 2
molecules in the same diamond network. Therefore, for perfect ice
(i.e., fixed arm configuration), the tetrahedral cluster
approximation describes sign-configuration statistics in the very
same way as the Bethe (2-site) approximation does.

According to these arguments, we restrict the minimization
procedure to a subspace of the probability distribution space,
characterized by equally probable sign configurations. It is
possible to show that such a calculation leads to a set of
equations for the probability distribution of arm configurations,
that are formally equivalent to Eqs.~\eqref{eq:cvmeq} and
\eqref{eq:costnorm}. In these equations, the tetrahedron
hamiltonian $\ham_{ijkl}$ can still be evaluated by
Eqs.~\eqref{eq:tetraham} and \eqref{eq:triham}, but the H bond
energy $\eta$ is replaced by the effective temperature-dependent
parameter
\begin{equation}
  \tilde{\eta}(T) = T \ln \frac{1+e^{\eta/T}}{2}
  ,
\end{equation}
while the chemical potential $\mu$ is replaced by
\begin{equation}
  \tilde{\mu}(T) = \mu + T \ln 6
  .
\end{equation}
(recall that we have 6 sign configurations for each arm
configuration). These parameters contain entropic contributions,
which take into account the extra degrees of freedom related to
the sign configurations. Such contributions must not be taken into
account in the evaluation of the (grand-canonical) average energy.
It turns out that the latter can still be computed by
Eqs.~\eqref{eq:intenergy}, \eqref{eq:tetraham}, and
\eqref{eq:triham}, with $\eta$ replaced by
\begin{equation}
  \bar{\eta}(T) = \eta \, \frac{e^{\eta/T}}{1+e^{\eta/T}}
  .
\end{equation}
The latter parameter represents the H-bond energy, averaged over
the sign configurations.

Let us finally recall that we are also interested in studying the
effect of a homogeneity constraint, i.e., of forcing site
probability distributions $p_i^X$ to be independent of the
sublattice $X$. It is easy to see that, if the latter condition is
verified in the right-hand side of Eq.~\eqref{eq:cvmeq} the
resulting tetrahedron distribution (left-hand side) turns out to
be invariant under circular permutation of the indices, due to the
same invariance property of the tetrahedron hamiltonian
$\ham_{ijkl}$. As a consequence, the marginals computed by
Eqs.~\eqref{eq:marginals} will verify the homogeneity condition as
well. This observation clearly shows that the cluster-variational
free energy actually admits ``homogeneous'' stationary points. In
order to determine such points, preventing numerical instabilities
which might drive the iterative procedure toward more stable
symmetry-broken solutions, it is sufficient to compute marginals
according to the following prescription
\begin{equation}
  p_i^A = p_i^B = p_i^C = p_i^D = \sum_{j,k,l} \frac{p_{ijkl}+p_{lijk}+p_{klij}+p_{jkli}}{4}
  ,
\end{equation}
rather than Eqs.~\eqref{eq:marginals}.

\section{Results}

\subsection{Phases}

Let us consider the probability distributions $p_i^X$ of the arm
configuration $i$ on the $X$ sublattice, as defined in
Eqs.~\eqref{eq:marginals}. Due to normalization, each distribution
can be represented by two parameters, which can be conveniently
defined as
\begin{eqnarray}
  \rho^X & \equiv & p_1^X + p_2^X , \\
  \xi^X & \equiv & p_1^X - p_2^X .
\end{eqnarray}
The former is an occupation probability, restricted to the $X$
sublattice, so that we may call it sublattice density. The latter
characterizes preference for the water arm configuration $1$
rather than $2$ on the $X$ sublattice. For the benefit of readers
that are familiar with spin models, let us note that the arm
configuration variables of the current models admit a ``natural''
mapping onto spin-1 variables, namely, $S=\pm 1$ spin values
representing the two water configurations, and $S=0$ representing
empty sites. In this view, the above parameters turn out to be the
usual order parameters of spin-1 models, i.e., the quadrupolar
order parameter $\langle S^2 \rangle$ and the magnetization
$\langle S \rangle$, respectively. It is possible to characterize
the different phases of our models by the set of sublattice order
parameters $\rho^X,\xi^X$, for $X=A,B,C,D$. Let us stress the fact
that the same characterization is valid for all the models under
investigation, both with symmetric and asymmetric bonds.

At high temperature and low pressure, we always observe a fully
homogeneous (gaslike or liquidlike) phase, which we simply denote
as fluid (F), where the order parameters $\rho^X,\xi^X$ are
independent of the sublattice. The $\xi^X$ parameters vanish,
revealing that there is no preference between the two alternative
arm configurations of a water molecule. In summary, we can write
\begin{equation}
  \left[ \begin{matrix}
  \rho^A & \rho^B & \rho^C & \rho^D \cr
  \xi^A & \xi^B & \xi^C & \xi^D \cr
  \end{matrix} \right] =
  \left[ \begin{matrix}
  \rho_\mathrm{F} & \rho_\mathrm{F} & \rho_\mathrm{F} & \rho_\mathrm{F} \cr
  0 & 0 & 0 & 0
  \end{matrix} \right]
  ,
  \label{eq:F_phase}
\end{equation}
where the matrix notation has been chosen for reasons of clarity.
In the low-temperature region, we observe two different phases,
which we denote as high density (HD) and low-density (LD) phases.
These phases can be regarded as a temperature evolution of the two
(HD and LD) ground-state ordered network structures, respectively.

As far as the HD phase is concerned, we have
\begin{equation}
  \left[ \begin{matrix}
  \rho^A & \rho^B & \rho^C & \rho^D \cr
  \xi^A  & \xi^B  & \xi^C  & \xi^D
  \end{matrix} \right] =
  \left[ \begin{matrix}
  \rho_\mathrm{HD}  & \rho_\mathrm{HD}  & \rho_\mathrm{HD}  & \rho_\mathrm{HD} \cr
  \xi_\mathrm{HD} & -\xi_\mathrm{HD} & \xi_\mathrm{HD} & -\xi_\mathrm{HD}
  \end{matrix} \right]
  ,
  \label{eq:HD_phase}
\end{equation}
with $\rho_\mathrm{HD}=\xi_\mathrm{HD}=1$ at zero temperature. The
density is still homogeneous, whereas the preference parameters
reveal that H bonds are more likely formed within the sublattice
pairs $AB$ and $CD$, respectively. In other words, this phase can
be viewed as an ideal HD structure (two interpenetrating diamond
networks), in which defects are progressively introduced by
thermal fluctuations. Like in the ground state, this phase turns
out to be twofold degenerate. The degenerate solution, having the
same free energy, can be obtained from Eq.~\eqref{eq:HD_phase} by
a circular permutation of the sublattice indices.

The LD phase is characterized by
\begin{equation}
  \left[ \begin{matrix}
  \rho^A & \rho^B & \rho^C & \rho^D \cr
  \xi^A  & \xi^B  & \xi^C  & \xi^D
  \end{matrix} \right] =
  \left[ \begin{matrix}
  \rho'_\mathrm{LD} &  \rho'_\mathrm{LD} & \rho''_\mathrm{LD} &  \rho''_\mathrm{LD} \cr
  \xi'_\mathrm{LD}  & -\xi'_\mathrm{LD}  & \xi''_\mathrm{LD}  & -\xi''_\mathrm{LD}
  \end{matrix} \right]
  .
  \label{eq:LD_phase}
\end{equation}
Assuming for instance $\rho'_\mathrm{LD}>\rho''_\mathrm{LD}$, the
sublattice pair $AB$ turns out to be more populated than $CD$,
whereas the preference parameters reveal the formation of
preferential bonding within the same sublattice pairs. This phase
can be viewed as a temperature evolution of the zero temperature
LD structure, characterized by
$\rho'_\mathrm{LD}=\xi'_\mathrm{LD}=1$ and
$\rho''_\mathrm{LD}=\xi''_\mathrm{LD}=0$. Thermal fluctuations
introduce defects in the diamond H bond network on the $AB$
sublattices, and begin to populate the $CD$ sublattices, which are
rigorously empty at zero temperature. This phase is fourfold
degenerate, and the alternative solutions can be obtained from
Eq.~\eqref{eq:LD_phase} by performing all possible circular
permutations.

In certain conditions, we also observe a peculiar phase, which we
denote as ``modulated'' fluid (MF) phase, characterized by the
following parameters
\begin{equation}
  \left[ \begin{matrix}
  \rho^A & \rho^B & \rho^C & \rho^D \cr
  \xi^A  & \xi^B  & \xi^C  & \xi^D
  \end{matrix} \right] =
  \left[ \begin{matrix}
  \rho'_\mathrm{MF} & \rho''_\mathrm{MF} & \rho'_\mathrm{MF} & \rho''_\mathrm{MF} \cr
  0 & 0 & 0 & 0
  \end{matrix} \right]
  .
  \label{eq:MF_phase}
\end{equation}
Assuming $\rho'_\mathrm{MF}>\rho''_\mathrm{MF}$, the two
sublattices $A$ and $C$ are more populated than $B$ and $D$
(whence the term ``modulated''), while there is no preference
between the two arm configurations (whence the term ``fluid'').
This phase can be viewed as a temperature evolution of a zero
temperature state with $\rho'_\mathrm{MF}=1$ (two fully populated
sublattices) and $\rho''_\mathrm{MF}=0$ (two empty sublattices),
or vice versa. Note that the two populated sublattices are next
nearest neighbors, so that there is no interaction energy in these
states. One can verify that these can actually be ground states of
the GBHB and BL models, only if the two-body repulsive interaction
is larger than the H bond energy, i.e., $\epsilon/\eta<-1$.

Let us finally remark that, except the fully homogeneous F phase,
all the phases described above (HD, LD, and MF) are
symmetry-broken phases, which do not possess the full
translational symmetry of the bcc lattice.

\subsection{GBHB model}

\label{subsec:results_GBHB}

\begin{figure}[t]
  \includegraphics*[40mm,153mm][125mm,245mm]{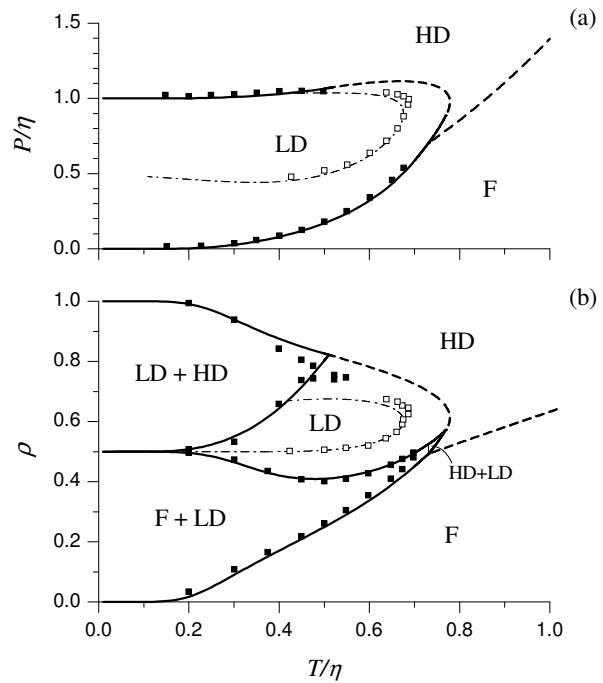}
  \caption
  {
    GBHB model phase diagram, $\epsilon/\eta=-0.5$:
    pressure vs temperature (a); density vs temperature (b).
    F, LD, HD denote the corresponding phases (see the text);
    double labels denote coexistence regions (b).
    Solid lines denote first-order transitions (a)
    or boundaries of coexistence regions (b);
    dashed lines denote second-order transitions;
    (thin) dash-dotted lines denote the TMD locus.
    Symbols display Monte Carlo data from
    Ref.~\onlinecite{GirardiBalladaresHenriquesBarbosa2007}:
    solid squares denote first-order transitions (a)
    or coexistence boundaries (b);
    open squares denote the TMD locus.
  }
  \label{fig:barbosa1}
\end{figure}
In Fig.~\ref{fig:barbosa1} we report the phase diagram of the GBHB
model, for $\epsilon/\eta=-0.5$. This ratio corresponds to the
parameters chosen by Girardi and coworkers in the original
paper~\cite{GirardiBalladaresHenriquesBarbosa2007}. In the $T$-$P$
diagram (Fig.~\ref{fig:barbosa1}a), we observe three different
first-order transition lines. The first one separates the F and LD
phases, whereas the other two occur between the LD and HD phases.
All these lines are mapped onto coexistence regions in the
$T$-$\rho$ diagram (Fig.~\ref{fig:barbosa1}b). Both the LD-HD
first-order transition lines terminate in tricritical points,
which are connected by a second-order line (i.e., a line of
critical points) enclosing the LD phase. Another critical line
separates the F phase from the HD phase, and terminates in a
critical end-point. The LD phase exhibits a density anomaly,
namely, a temperature of maximum (or minimum) density (TMD), at
constant pressure. In the $T$-$P$ diagram, the TMD locus is
bounded between a minimum and a maximum pressure. The portion of
the TMD line joining these two points corresponds to density
maxima, whereas the remaining two branches of the line to density
minima.

Fig.~\ref{fig:barbosa1} also reports some data points obtained by
the Monte Carlo study of
Ref.~\cite{GirardiBalladaresHenriquesBarbosa2007}. We find a
remarkably good agreement both for first-order transitions
(transition lines and coexistence regions in the $T$-$P$ and
$T$-$\rho$ diagrams, respectively), and for the TMD locus. Such an
agreement suggests that the tetrahedral cluster approximation is
able to take into account the most relevant correlations present
in the system. Let us recall that in
Ref.~\onlinecite{GirardiBalladaresHenriquesBarbosa2007} the
authors could not detect second-order transitions, and interpreted
the tricritical points as critical points terminating first-order
transitions between homogeneous fluid phases of different
densities. The misinterpretation was probably due to the fact that
no order parameter was investigated. Indeed, in
Ref.~\onlinecite{BuzanoDeStefanisPretti2008} we collected several
evidences supporting the existence of the ordered phases predicted
by the cluster approximation, and the validity of the
corresponding phase diagram.

The previous results show that unfortunately the phase diagram of
the GBHB model is quite far from that of real water. The main
difficulty is that the density anomaly appears in a region where
the system still exhibits orientational order (i.e., the icelike
LD phase). Furthermore, the LD phase is never less dense than the
corresponding fluid phase, as shown by the slope of the F-LD
transition line in the $T$-$P$ diagram, and no liquid-vapor
coexistence is observed. We have explored the possibility of
removing at least the first difficulty, by reducing the relative
importance of the attractive H-bond interaction (which is, the one
responsible for orientational order), with respect to the
repulsive close-packing interaction.

\begin{figure}[t]
  \includegraphics*[40mm,143mm][125mm,245mm]{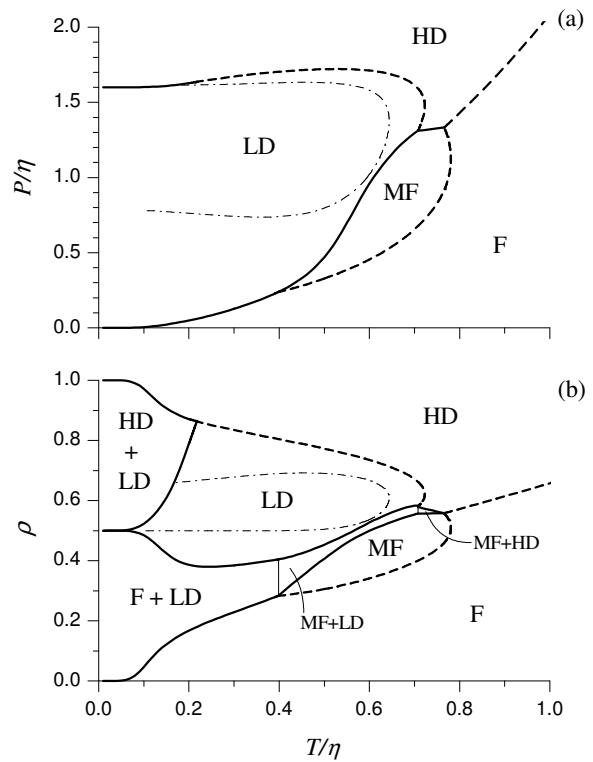}
  \caption
  {
    GBHB model phase diagram, $\epsilon/\eta=-0.8$:
    pressure vs temperature (a); density vs temperature (b).
    Labels and lines are defined as in Fig.~\ref{fig:barbosa1};
    MF denotes the modulated fluid phase.
  }
  \label{fig:barbosa2}
\end{figure}
In Fig.~\ref{fig:barbosa2} we report the phase diagram for
$\epsilon/\eta=-0.8$. It turns out that the stability of the LD
phase with respect to temperature is actually reduced, but our
difficulty is not solved, since the TMD locus is displaced toward
lower temperatures, remaining inside the LD phase region.
Furthermore, a modulated fluid (MF) phase becomes stable at finite
temperature, which makes the phase diagram even more complex. The
MF phase is separated from the ordinary fluid (F) phase by a
second-order transition and from the LD and HD phases by two
different first-order transitions. The physical mechanism
underlying the onset of the MF phase can be qualitatively
understood by energetic arguments. On the one hand, the increased
importance of nearest-neighbor repulsion favors configurations in
which occupied sites have empty neighbors (by the way, this also
enhances stability of the LD phase with respect to pressure). On
the other hand, the reduced importance of H bonding favors
thermodynamic states with no preference for particular molecule
orientations.

We have also investigated the opposite regime, in which the H-bond
interaction dominates with respect to the close-packing
interaction (i.e., $\eta \gg |\epsilon|$). We have observed that
the pressure and density ranges on which the LD phase is stable
become smaller and smaller, but the phase diagram does not exhibit
significant changes with respect to the case $\epsilon/\eta=-0.5$.

\subsection{BL model}

\label{subsec:results_BL}

\begin{figure}[t]
  \includegraphics*[40mm,153mm][125mm,245mm]{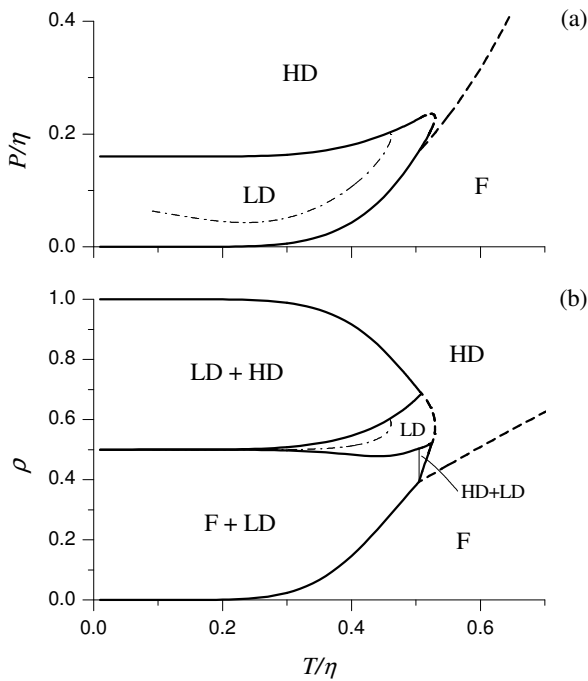}
  \caption
  {
    BL model phase diagram, $\epsilon/\eta=-0.08$:
    pressure vs temperature (a); density vs temperature (b).
    Labels and lines are defined as in Fig.~\ref{fig:barbosa1}.
  }
  \label{fig:besseling1}
\end{figure}

As mentioned in the Introduction, the BL model can be viewed as a
version of the GBHB model, with asymmetric bonds. The interaction
parameters chosen by BL in the original
paper~\cite{BesselingLyklema1994}, based on a fit to real
thermodynamic properties, correspond to $\epsilon/\eta=-0.08$,
which is, to a regime of dominating H-bond interaction. The phase
diagram we have obtained for this case is reported in
Fig.~\ref{fig:besseling1}.
This phase diagram is, both qualitatively and quantitatively, very
similar to the one of the GBHB model for the same $\epsilon/\eta$
value. We just observe a sort of ``rescaling'' of the whole phase
diagram toward lower temperatures. This effect is of entropic
nature, and is related to the presence of the sign configurations.

\begin{figure}[t]
  \includegraphics*[40mm,153mm][125mm,245mm]{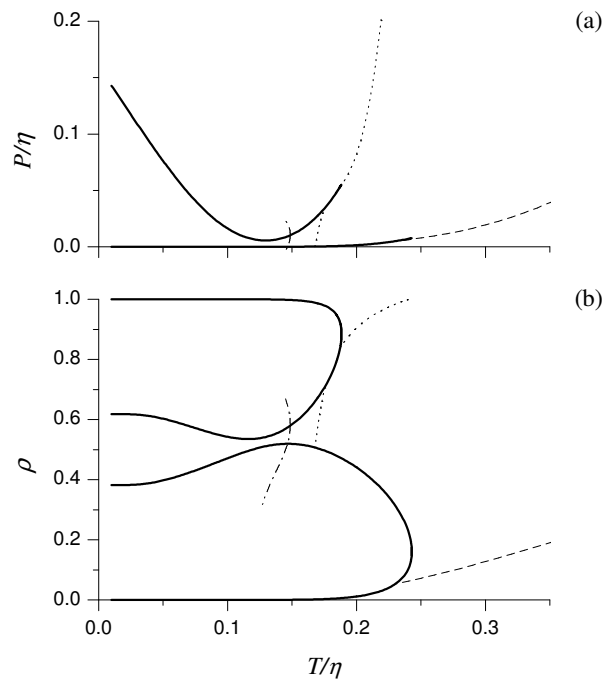}
  \caption
  {
    BL model phase diagram in the homogeneity assumption, $\epsilon/\eta=-0.08$:
    pressure vs temperature (a); density vs temperature (b).
    Dashed and dotted lines denote respectively the
    stability limit and the zero-entropy locus.
    Other lines are defined as in Fig.~\ref{fig:barbosa1}.
  }
  \label{fig:besseling2}
\end{figure}

Apart from these changes, the most remarkable fact in the above
phase diagram is that we find no trace of the vapor-liquid
coexistence, claimed by BL~\cite{BesselingLyklema1994}. We have
already mentioned that the reason of such a discrepancy is a
homogeneity hypothesis, on which the calculations of the cited
paper are based. As shown in the previous Section, our approach
allows one to impose an analogous homogeneity constraint in a
straightforward way. We have repeated the phase diagram
calculation, taking into account this constraint, and the results
are reported in Fig.~\ref{fig:besseling2}.
Of course, we no longer observe the ordered phases, but only
homogeneous (fluid) phases. At low temperature, there appear two
first-order transition lines (and related coexistence regions),
allowing us to distinguish three phases with different densities.
In particular, the transition between the lowest density phase and
the intermediate density one might be identified with the
vapor-liquid transition. The other transition, between the
intermediate and the highest density phases, reminds us of the
conjecture about the liquid-liquid transition in metastable water,
put forward by Stanley and
coworkers~\cite{PooleSciortinoEssmannStanley1992}. Unfortunately,
a more detailed analysis shows that this ``homogeneous'' phase
diagram carries several inconsistencies. First of all, one can
observe that the entropy is negative below a certain temperature,
which depends on pressure. The TMD locus and most part of the
supposed liquid-liquid coexistence region (including the critical
point), lie below the zero-entropy line. Moreover, making use of
Eq.~\eqref{eq:hessian}, it is possible to compute the continuation
of the F-HD critical line in the stability region of the LD phase.
This line marks a boundary, beyond which the free-energy minimum
associated to the homogeneous solution becomes a saddle point,
i.e., the homogeneous phase becomes thermodynamically unstable. As
shown in Fig.~\ref{fig:besseling2}, such a condition comes true in
a very large portion of the phase diagram. By the way, let us also
remark that, although for brevity we do not report the results,
totally analogous phase diagrams have been obtained for the
``homogeneous'' GBHB model.

We have also reproduced BL's original
calculation~\cite{BesselingLyklema1994}, in order to compare the
results with those obtained by our approach, in the homogeneity
assumption.
\begin{figure}[t]
  \includegraphics*[40mm,143mm][125mm,245mm]{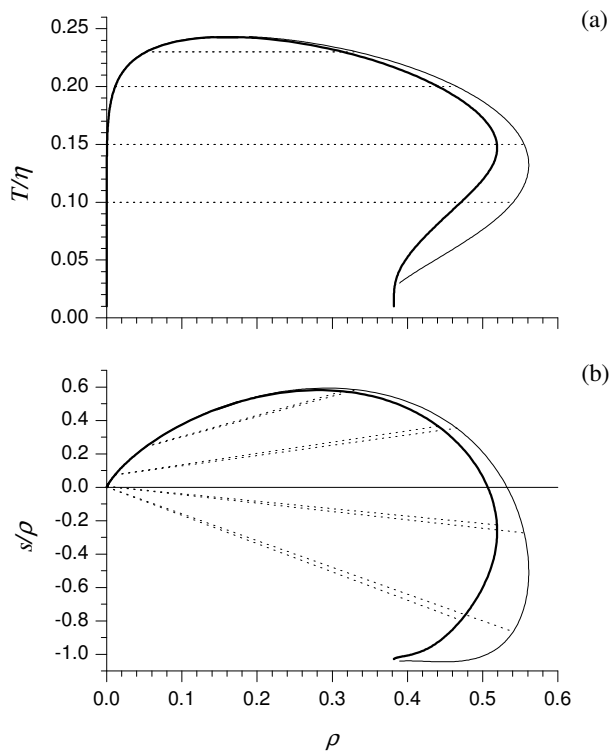}
  \caption
  {
    Liquid-vapor coexistence region for the BL model in the
    homogeneity assumption, $\epsilon/\eta=-0.08$:
    temperature vs density (a); entropy per molecule vs density (b).
    Thicker and thinner lines refer to our tetrahedral cluster approximation
    and BL's pair approximation, respectively.
    Solid lines denote the binodals;
    straight dotted lines denote coexistence lines at different temperatures.
  }
  \label{fig:besseling3}
\end{figure}
Fig.~\ref{fig:besseling3} reports the vapor-liquid binodals and
some coexistence lines in the density-temperature and
density-entropy planes. It turns out that, for not too high
densities, the two methods yield very similar results. Conversely,
a relevant discrepancy appear in the high density regime, which
is, the phase diagram computed by BL~\cite{BesselingLyklema1994}
does not exhibit the ``liquid-liquid'' transition at all. Such a
discrepancy is ultimately due to the different cluster considered
for the approximation, namely, a nearest-neighbor pair in BL's
calculation and the tetrahedral cluster in the current one. The
(pseudo) liquid-liquid transition is indeed a reminiscence of the
distinction between the LD and HD structures, which the
tetrahedron approximation is able to account for (although, in the
homogeneous assumption, only at the level of local correlations),
at odd with the pair approximation.

Let us finally remark that, according to
Fig.~\ref{fig:besseling3}b, even BL's solution turns out to suffer
from the negative-entropy problem. More generally, one can see
that all the aforementioned inconsistencies do not depend on the
type of approximation (except the fact that the pair approximation
does not reveal the liquid-liquid transition) but rather on the
artificial homogeneity constraint.

\subsection{Bell model}

\begin{figure}[t]
  \includegraphics*[40mm,143mm][125mm,245mm]{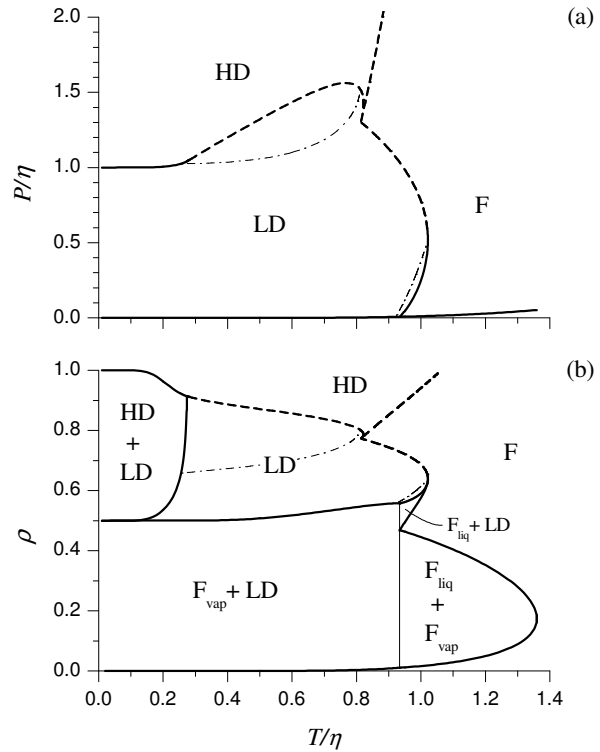}
  \caption
  {
    Bell model phase diagram, $\epsilon/\eta=2$, $3\gamma/\eta=-5/4$:
    pressure vs temperature (a); density vs temperature (b).
    Labels and lines are defined as in Fig.~\ref{fig:barbosa1}.
    Different fluid phases with lower and higher densities are
    denoted by $\mathrm{F_{vap}}$ and $\mathrm{F_{liq}}$, respectively.
  }
  \label{fig:bell1}
\end{figure}

The Bell model is quite different from the GBHB and BL models, as
the nearest-neighbor interaction is attractive, while the
close-packing repulsion effect is modeled by a three-body
interaction (see Section II). For the parameter set of the
original paper~\cite{Bell1972}, we have obtained the phase diagram
reported in Fig.~\ref{fig:bell1}.
We can indeed observe several differences, with respect to the
phase diagrams presented above. First of all, there appears a
coexistence region between two homogeneous (fluid) phases at
different densities, which we can identify with a vapor and a
liquid phase, respectively. This new feature is likely to be
related to the quite strong orientation-independent attractive
interaction ($\epsilon>0$), which is absent in the other models.
The ordered LD and HD phases are still present, but with clearly
different phase transitions. The F phase turns out to be more
stable with respect to pressure, so that a direct F-LD transition
takes place, at odd with the GBHB and BL models. Such a transition
is partially first-order (at lower pressure) and partially
second-order (at higher pressure). As a consequence, a
multicritical point appears, at which three critical lines (F-LD,
LD-HD, and F-HD) merge. Density anomalies can still be observed in
the LD phase. The TMD locus is made up of two different branches:
a lower pressure one, corresponding to density maxima, and a
higher pressure one, corresponding to density minima. The F phase
does not display any density anomaly.

\begin{figure}[t]
  \includegraphics*[40mm,153mm][125mm,245mm]{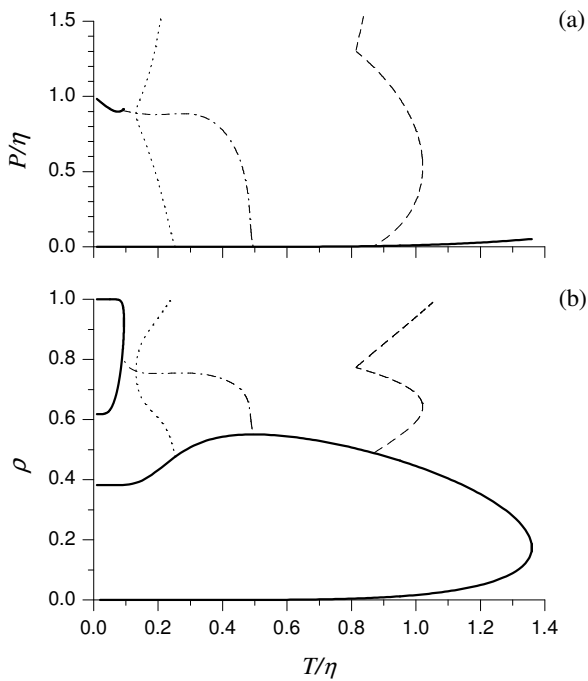}
  \caption
  {
    Bell model phase diagram in the homogeneity assumption,
    $\epsilon/\eta=2$, $3\gamma/\eta=-5/4$:
    pressure vs temperature (a); density vs temperature (b).
    Lines are defined as in Fig.~\ref{fig:besseling2}.
  }
  \label{fig:bell2}
\end{figure}

Let us now consider the phase diagram obtained imposing the
homogeneity constraint (Fig.~\ref{fig:bell2}).
Of course, no ordered phase is present, whereas the vapor-liquid
coexistence region continues down to zero temperature. Let us note
that Bell's original calculation~\cite{Bell1972} is equivalent to
the tetrahedral cluster approximation for homogeneous phases, so
that the vapor-liquid binodals, obtained by the former, turn out
to be exactly superimposed to those reported in
Fig.~\ref{fig:bell2}. In the very low temperature region, we also
observe a ``liquid-liquid'' coexistence, like in the BL and GBHB
models. The latter feature was not pointed out by
Bell~\cite{Bell1972}, most likely because at that time the
exploration of supercooled water physics was at the very
beginning~\cite{RasmussenMacKenzieAngellTucker1973}, and the
``second critical point'' conjecture had not been proposed
yet~\cite{PooleSciortinoEssmannStanley1992}. Anyway, the entropy
computed by the cluster approximation turns out to be negative in
this region. Conversely, the TMD locus lies mostly in the positive
entropy region, and exhibits the correct experimental slope. The
latter was indeed one of the most striking results of the Bell
model. Unfortunately, the stability limit, which in this case is
made up of both the F-HD and F-LD second-order transition lines
(and a metastable continuation of the latter), suggests that, even
for the Bell model, all the interesting anomalies take place in a
region where the homogeneous F phase is thermodynamically
unstable.

We have compared our results for the Bell model with some Monte
Carlo data obtained by Whitehouse and
coworkers~\cite{WhitehouseChristouNicholsonParsonage1984}.
\begin{figure}[t]
  \includegraphics*[40mm,191mm][125mm,245mm]{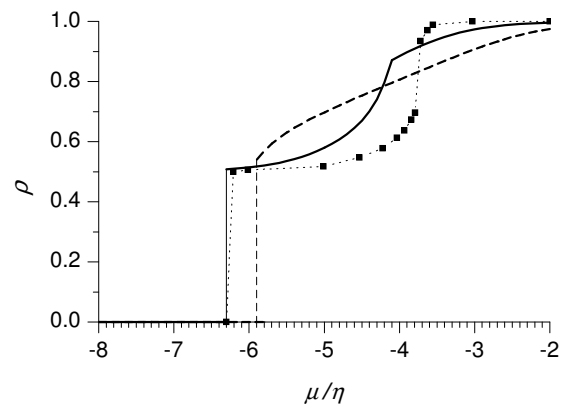}
  \caption
  {
    Density vs chemical potential at constant temperature $T/\eta=0.5$
    for the Bell model, $\epsilon/\eta=2$, $3\gamma/\eta=-5/4$.
    The dashed and solid lines denote the predictions of
    the cluster-variational calculation, respectively with or without
    the homogeneity constraint. Symbols denote Monte Carlo results
    from Ref.~\onlinecite{WhitehouseChristouNicholsonParsonage1984}
    (the thin dotted line is an eye-guide).
  }
  \label{fig:bell3}
\end{figure}
In Fig.~\ref{fig:bell3} we report the density as a function of the
chemical potential at fixed temperature $T/\eta=0.5$. At a first
glance, it turns out that the tetrahedral cluster approximation
does not fit the Monte Carlo results so closely as for the GBHB
model. This fact is probably to be ascribed to stronger
correlations, present in the Bell model, arising from the
three-site interaction. We expect that the observed discrepancy is
not so relevant to affect the qualitative structure of the phase
diagram. Anyway, it is evident that the symmetry-broken solution
(solid line) is much closer to the Monte Carlo results than the
homogeneous one (dashed line). In particular, the LD-HD symmetry
change (see Subsection A) provides an explanation for the
transition observed at higher chemical potential. Therefore, we
argue that the simulations of Whitehouse and
coworkers~\cite{WhitehouseChristouNicholsonParsonage1984} actually
describe ordered phases of the same type predicted by our
approximation. This fact was not recognized in the cited paper,
most likely because the authors did not investigate any kind of
order parameter.

\section{Conclusions}

The basic ideas of the present paper originated in a previous
one~\cite{BuzanoDeStefanisPretti2008}, in which we showed that a
generalized first-order approximation on a tetrahedral cluster was
able to reproduce several Monte Carlo results for a waterlike
network-forming lattice model (GBHB
model)~\cite{GirardiBalladaresHenriquesBarbosa2007}. The
approximate theory also predicted that the two denser phases,
identified as liquid in
Ref.~\onlinecite{GirardiBalladaresHenriquesBarbosa2007}, were in
fact characterized by two different ordered H-bond-network
structures. In the current work, we have exploited the same
cluster approximation, to extend the analysis of the GBHB model to
different parameter values and to revisit other models with
similar features (tetrahedral molecules, bcc lattice), already
known in the literature (BL and Bell models).

We find that analogous ordered phases are present in all the
models investigated. As mentioned in the text, several previous
investigations did not point out these phases. In particular,
while Girardi and coworkers did not recognize order-disorder
transitions in their
simulations~\cite{GirardiBalladaresHenriquesBarbosa2007}, as they
did not compute order parameters, other studies ignored the
ordered phases by imposing homogeneity in an analytical
way~\cite{Bell1972,BesselingLyklema1994,BesselingLyklema1997}. Our
results show that the onset of orientational order produces
substantial changes in the phase diagrams, which turn out to be
very complex and rich, but quite far from real water.

Let us recall that, in principle, the ordered phases are an
interesting feature of the current models, since they might be
regarded as a description of two different ice forms.
Unfortunately, such an interpretation gives rise to some
difficulties. The most relevant one is that the icelike LD phase
exhibits a temperature of maximum density, which would be indeed
expected in the liquidlike F phase. Conversely, the F phase turns
out to be rather trivial, as it does not exhibit any density
anomaly, and, in the GBHB and BL models, not even a vapor-liquid
transition. These difficulties seem to be somehow related to the
fact that, in these models, orientational order is exceedingly
stable with respect to thermal fluctuations.

We have explored the possibility that a suitable choice of the
parameter set could yield a more realistic phase diagram, in the
framework of the same models. Several failed attempts have led us
to believe that this is likely not to be the case. Indeed, the
physical mechanism underlying density anomalies (which is,
correlation between low energy and high specific-volume
configurations) is also responsible for enhancing thermodynamic
stability of the ordered LD phase. Accordingly, a parameter change
oriented, for instance, to lowering the LD-F transition
temperature, usually turns out to lower the temperature of maximum
density as well. In Section~\ref{subsec:results_GBHB} we have
reported an example of such an effect for the GBHB model.

Let us remark that analogous arguments might probably hold for
other similar models, not studied in this paper. For instance,
Bell and Salt~\cite{BellSalt1976} have shown that their model
predicts open- and close-packed icelike phases, which are
respectively equivalent to the LD and HD phases of the current
models. More recently, it was
noticed~\cite{RobertsPanagiotopoulosDebenedetti1996} that the
liquidlike phase of the Roberts-Debenedetti model is metastable at
``ordinary'' temperatures. Indeed, the authors of
Ref.~\onlinecite{RobertsPanagiotopoulosDebenedetti1996} devise
particular techniques to prevent the Monte Carlo dynamics from
``falling'' into ordered states. Two authors of the present paper
have verified the onset of analogous ordered states in a
simplified (symmetric bonds) version~\cite{PrettiBuzano2004} of
the Roberts-Debenedetti model, although this result has not been
published yet. All these facts suggest that a quite large class of
waterlike lattice models should be critically reconsidered.

Concerning the stability of the H-bond network structures, let us
note that, in the models under investigation, this is necessarily
overemphasized by the regular bcc lattice, which imposes a strong
directional correlation on the bonds. Such a correlation is not
present in the real (off-lattice) system. In particular, it is
known from experiments that directional correlation in real water
is almost completely lost after the second consecutive bond (see
Ref.~\onlinecite{CabaneVuilleumier2005} and references therein).
Therefore, a more realistic model might be provided by a suitable
random lattice, preserving the local geometric structure of the
H-bond network (up to the second consecutive bond). We expect that
such a lattice might hinder the onset of orientational order,
though retaining the essential physical properties of the
liquidlike phase. Of course, even a random-lattice description
involves a simplification, because, in principle, directional
disorder should not be an a-priori assumption, but rather a
prediction of the theory.

In any case, a very simple realization of the random lattice,
satisfying the requirement on the local tetrahedral geometry, is
the Husimi lattice made up of tetrahedral blocks, which we have
mentioned in Section IV (Fig.~\ref{fig:cactus}). As noted there,
the tetrahedral cluster approximation happens to be exact for this
particular lattice. Furthermore, it is known that, in general,
Husimi and Bethe lattices are characterized by a random graph
structure, which is locally treelike, but contains closed loops of
different (even and odd) lengths. The latter fact can produce
frustration, which is expected to forbid periodically ordered
states~\cite{RivoireBiroliMartinMezard2004} like those appearing
on a regular lattice. As a consequence, it turns out that the
different models treated in this paper, properly redefined on the
Husimi lattice, cannot exhibit ordered phases at all. The exact
thermodynamic equilibrium states for these models are given by the
``homogeneous'' stationary points of the appropriate variational
free energy. This argument allows one to reinterpret the
``homogeneous'' phase diagrams, presented in this paper, which
exhibit interesting similarities with real water thermodynamics.
In the new interpretation, such results are no longer artifacts of
the homogeneity assumption, but rather exact solutions for the
corresponding Husimi lattice models.

At first sight, the above idea might look a bit tricky, but in
fact a random lattice is not much more arbitrary than any regular
lattice, for modeling a fluid. Both types of lattice should be
regarded as simplifying assumptions, which affect the model
predictions in opposite ways. In the regular lattice case, the
stability of the ordered phases is overestimated, whereas, in a
generic random lattice, it is underestimated. In the simplest case
of a Husimi lattice, whose exact solution is almost analytically
available, periodically ordered states turn out to be rigorously
forbidden. In other words, the price paid for simplicity is that
the model is no longer able to describe icelike phases. Let us
also remind that a Husimi lattice is an infinite-dimensional
system, so that all its properties have necessarily a mean-field
nature. As a consequence, the model cannot yield realistic
estimates of critical exponents, which turn out to belong to the
mean-field universality class.

Let us also stress the fact that the tetrahedral cluster is not
the only possible choice for the building block of the Husimi
lattice. This choice turns out to be quite effective for the
current models, as it is able to distinguish two different (low-
and high-density) local molecular packings. As mentioned in
Section~\ref{subsec:results_BL}, these two packings are not only
the elementary structures of the ideal LD and HD networks (which
are not relevant on the random lattice), but they are also
expected to play a role in the possible onset of a liquid-liquid
phase transition.

Let us finally note that, in the light of the new interpretation,
the negative-entropy phenomenon suggests that the waterlike Husimi
lattice models might predict a replica symmetry breaking, in
analogy with the models studied by M\'ezard and
coworkers~\cite{RivoireBiroliMartinMezard2004,BiroliMezard2002}.
This possibility might even be relevant for describing the glassy
states of water (amorphous ices), which have been studied
intensively in the last years by both experiments and
simulations~\cite{Debenedetti2003}. We are going to investigate
this issue in a future work.


\end{document}